\documentclass[twocolumn]{aastex631}
\usepackage{CJK}
\usepackage{amsmath}

\begin{document}
\begin{CJK*}{UTF8}{gbsn}

\title{Ly$\alpha$ imaging around the hyperluminous dust-obscured quasar W2246$-$0526 at $z=4.6$ }

\correspondingauthor{Yibin Luo,Lulu Fan}
\email{yibinluo@mail.ustc.edu.cn,llfan@ustc.edu.cn}

\author[0000-0002-8079-6525]{Yibin Luo （罗毅彬）}
\affiliation{CAS Key Laboratory for Research in Galaxies and Cosmology, Department of Astronomy, University of Science and Technology of China, Hefei 230026, China}

\affil{School of Astronomy and Space Science, University of Science and Technology of China, Hefei 230026, China}

\author[0000-0003-4200-4432]{Lulu Fan (范璐璐)}
\affiliation{CAS Key Laboratory for Research in Galaxies and Cosmology, Department of Astronomy, University of Science and Technology of China, Hefei 230026, China}

\affil{School of Astronomy and Space Science, University of Science and Technology of China, Hefei 230026, China}

\affil{Deep Space Exploration Laboratory, Hefei 230088, China}

\author[0000-0002-2725-302X]{Yongming Liang （梁永明）}
\affil{Institute for Cosmic Ray Research, The University of Tokyo, 5-1-5 Kashiwanoha, Kashiwa, Chiba 277-8582, Japan}

\author[0000-0003-3424-3230]{Weida Hu （胡维达）}
\affiliation{Department of Physics and Astronomy, Texas A\&M University, College Station, TX, 77843-4242 USA}
\affiliation{George P.\ and Cynthia Woods Mitchell Institute for
 Fundamental Physics and Astronomy, Texas A\&M University, College Station, TX, 77843-4242 USA}

\author[0000-0002-4419-6434]{Junxian Wang （王俊贤）}
\affiliation{CAS Key Laboratory for Research in Galaxies and Cosmology, Department of Astronomy, University of Science and Technology of China, Hefei 230026, China}

\affil{School of Astronomy and Space Science, University of Science and Technology of China, Hefei 230026, China}

\author[0000-0002-9634-2923]{Zhen-ya Zheng （郑振亚）}
\affil{CAS Key Laboratory for Research in Galaxies and Cosmology, Shanghai Astronomical Observatory, Shanghai 200030, China}

\author[0000-0003-4959-1625]{Zheyu Lin （林哲宇）}
\affiliation{CAS Key Laboratory for Research in Galaxies and Cosmology, Department of Astronomy, University of Science and Technology of China, Hefei 230026, China}

\affil{School of Astronomy and Space Science, University of Science and Technology of China, Hefei 230026, China}

\author{Bojun Tao （陶柏钧）}
\affiliation{CAS Key Laboratory for Research in Galaxies and Cosmology, Department of Astronomy, University of Science and Technology of China, Hefei 230026, China}

\affil{School of Astronomy and Space Science, University of Science and Technology of China, Hefei 230026, China}

\author[0000-0001-8078-3428]{Zesen Lin （林泽森）}
\affiliation{Department of Physics, The Chinese University of Hong Kong, Shatin, N.T., Hong Kong S.A.R., China}

\author{minxuan Cai}
\affiliation{CAS Key Laboratory for Research in Galaxies and Cosmology, Department of Astronomy, University of Science and Technology of China, Hefei 230026, China}

\affil{School of Astronomy and Space Science, University of Science and Technology of China, Hefei 230026, China}

\author[0009-0003-5280-0755]{mengqiu Huang （黄梦秋）}
\affiliation{CAS Key Laboratory for Research in Galaxies and Cosmology, Department of Astronomy, University of Science and Technology of China, Hefei 230026, China}

\affil{School of Astronomy and Space Science, University of Science and Technology of China, Hefei 230026, China}

\author[0000-0002-3105-3821]{Zhen Wan （宛振）}
\affiliation{CAS Key Laboratory for Research in Galaxies and Cosmology, Department of Astronomy, University of Science and Technology of China, Hefei 230026, China}

\affil{School of Astronomy and Space Science, University of Science and Technology of China, Hefei 230026, China}

\author{Yongling Tang （唐永灵）}
\affiliation{CAS Key Laboratory for Research in Galaxies and Cosmology, Department of Astronomy, University of Science and Technology of China, Hefei 230026, China}

\affil{School of Astronomy and Space Science, University of Science and Technology of China, Hefei 230026, China}

\begin{abstract}
Hot dust-obscured galaxies (Hot DOGs) are a population of hyperluminous, heavily obscured quasars discovered by the \emph{Wide-field Infrared Survey Explorer} (\emph{WISE}) all-sky survey at high redshift. Observations suggested the growth of these galaxies may be driven by mergers. Previous environmental studies have statistically shown Hot DOGs may reside in dense regions. Here we use the Very Large Telescope (VLT) narrowband and broadband imaging to search for Ly$\alpha$ emitters (LAEs) in the $6.8\arcmin \times 6.8\arcmin$ field of the Hot DOG  W2246$-$0526 at $z=4.6$.  W2246$-$0526 is the most distant Hot DOG. We find that there is an overdensity of LAEs in  W2246$-$0526 field compared with the blank fields. This is the direct evidence that this most distant Hot DOG is in an overdense environment on the Mpc scale, and the result relates to the merger origin of Hot DOGs.
\end{abstract}


\section{Introduction} \label{sec:intro}

Based on ``W1W2-dropout" method, a new population of hyperluminous, hot dust-obscured galaxies were discovered using the WISE and were called as Hot DOGs \citep{Eisenhardt2012,Wu2012}. These galaxies are prominent in the WISE 12 $\mu m$ (W3) and 22 $\mu m$ (W4) bands, but are very faint or undetected in the 3.4 $\mu m$ (W1) and 4.6 $\mu m$ (W2) bands. Previous studies have found that Hot DOGs are extremely luminous $L_{bol} > 10^{13}L_\odot$, heavily dust-obscured quasars at high redshift, and represent a transition phase between starburst-dominated phase and optically bright quasars phase \citep{Assef2015,Fan2016a,Fan2018,Fan2020,Piconcelli2015,Stern2014,Sun2024,Aranda2024}.

Models predict that the growth of these galaxies may be driven by mergers \citep{Hopkins2006,Matteo2008}. Galaxy mergers remove gas angular momentum and drive it directly into the center, fueling intense starbursts and central SMBH accretion. Multiwavelength observations of Hot DOGs also suggested that Hot DOGs are likely triggered by galaxy mergers. \citet{Fan2016b} found a high merger fraction ($62\pm 14\%$) in a sample of 18 Hot DOGs using Hubble Space Telescope (HST) WFC3 imaging. Atacama Large Millimeter Array (ALMA) observations of a sample of 7 Hot DOGs suggested that there may be multiple merger events at the stage of Hot DOGs \citep{Diaz2021}. High merger fraction may be associated with high-density regions. Although some statistical studies have found overdensities of mid-IR-selected and sub-millimeter-selected galaxies around Hot DOGs \citep{Assef2015,Jones2014,Jones2017,Fan2017}, and two case studies have found overdensities of distant red galaxies (DRGs) \citep{Luo2022} and Lyman Break Galaxies (LBGs) \citep{Zewdie2023} around Hot DOGs, direct observations such as spectroscopically confirmed companion galaxies or narrowband observations are still very rare. \citet{Ginolfi2022} found an overdensity of Lyman-alpha emitters (LAEs) around a $z = 3.6$ Hot DOG using the VLT/MUSE. \citet{Diaz2018} found a $z = 4.6$ Hot DOG (W2246$-$0526, which is the most distant Hot DOG) in a multiple merger system using ALMA observations, with at least three spectroscopically confirmed companion galaxies within a few tens of kpc. Due to the limitations of the field of view of ALMA, the study of environment was restricted to scales of a few tens of kpc. Considering previous environment studies suggested that Hot DOGs may be a good tracer for overdense regions such as protoclusters, it is important to further study the environments of Hot DOGs in clusters scales of Mpc.

A commonly used technique to identify high-redshift galaxies is to search for sources with a prominent Ly$\alpha$ emission using specific narrowband filters ($\Delta \lambda \sim 100$\AA). These sources are called as Ly$\alpha$ emitters (LAEs). The covered redshift range of this technique is narrow ($\Delta z\sim 0.1$), which can reduce the impact of projection effects on environment studies. Numerous studies have used this technique to study the large scale environment \citep{Venemans2002,Venemans2005,Venemans2007,Cai2017,Mazzucchelli2017,Kikuta2017,Zheng2017,Garc2019,Hu2019,Liang2021,Hu2021}.

In this work, we use narrowband and broadband images obtained from VLT FOcal Reducer and low dispersion Spectrograph 2 (FORS2, \citealp{appenzeller1992}) to study the environment of the Hot DOG W2246$-$0526 (hereafter, W2246) by searching LAEs in this field. The redshift of W2246 derived from ALMA {\sc [Cii]} line is $z=4.601$ \citep{Diaz2016}. Among the Hot DOGs with spectroscopic redshift, W2246 is the most distant one so far. In addition, W2246 is the most luminous one with bolometric luminosity $L_{bol} > 10^{14}L_\odot$ \citep{Tsai2015}. Strong AGN feedback was found in W2246 \citep{Diaz2016}. \citet{Fan2018} found that most of its IR luminosity come from AGN torus, suggesting the rapid growth of the central SMBH. \citet{Tsai2018} measured the SMBH mass of about $10^{10} M_\odot$, and the Eddington ratio of 2.8.

The paper is structured as follows. We present the observation and data reduction in Section \ref{sec:obs}. We describe the color criteria and LAE sample in Section \ref{sec:selection}. Results and discussions are described in Section \ref{sec:results} and Section \ref{sec:discussion}, respectively. Finally, we give the summary and conclusion in Section \ref{sec:summary}.
Throughout this work, we use the AB magnitude system and assume a cosmology with $H_0=\rm{70\ km\ s^{-1}\ Mpc^{-1}}$, $\Omega_M = 0.27$ and $\Omega_\Lambda = 0.73$ \citep{komatsu2011}. All magnitudes are corrected for Galactic extinction \citep{SF2011}.

\begin{figure*}
\plotone{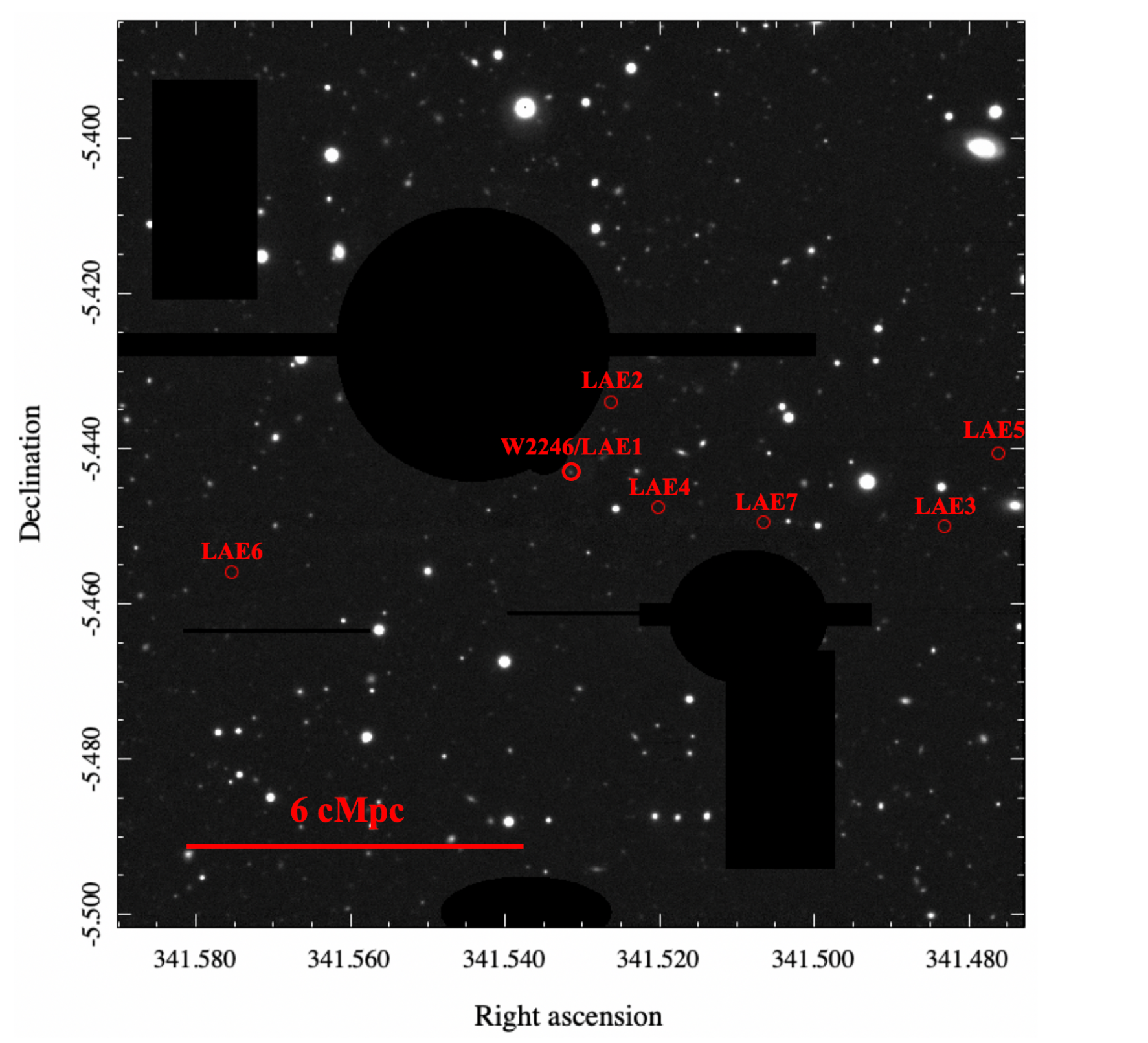}
\caption{$NB$ image of the W2246 field and the effective area is 33.6 arcmin$^2$. Positions of the LAEs are shown using the red circle and the bold red circle marks the position of W2246.
\label{fig1}}
\end{figure*}

\section{Observation and Data Reduction} \label{sec:obs}

We obtained narrowband and broadband imaging of the W2246 field with the FORS2 at the VLT in 2017 June and July. We used the red sensitive detector consisting of two 2k$\times$4k MIT CCDs and adopted a 2$\times$2 binning readout mode. The pixel scale is 0\arcsec.25/pixel, and the field of view (FOV) of FORS2 is 6.8$\times$6.8 arcmin$^2$, corresponding to 15.45$\times$15.45 $\rm{cMpc^{2}}$ at $z=4.6$. Here $\rm{cMpc}$ is the abbreviation of comoving $\rm{Mpc}$. The field was observed in the narrowband filter SII/2000+63 ($\lambda_c = 6774$\AA, FWHM = $68$\AA, hereafter $NB$) and the broadband filters R\_SPECIAL ($\lambda_c = 6550$\AA, FWHM = $1650$\AA, hereafter $R$) and V\_HIGH ($\lambda_c = 5550$\AA, FWHM = $1232$\AA, hereafter $V$). The total exposure times for $NB$, $R$, and $V$ bands are about 5.3, 1.0 and 1.5 hours, respectively.

\begin{figure}
\plotone{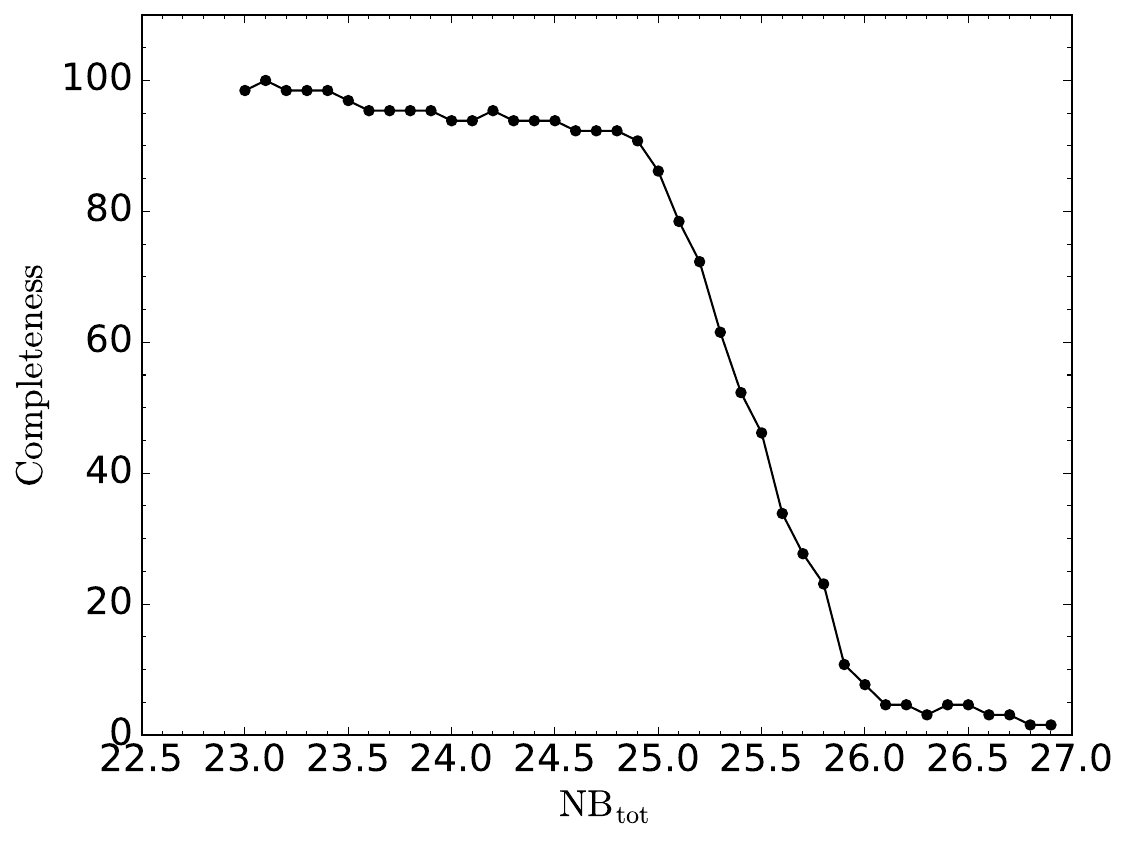}
\caption{
The detection completeness of the $NB$ image as a function of $NB_{tot}$ magnitude. The detection completeness drops to about 80 percent at $NB_{tot} = 25.1$ mag.
\label{fig2}}
\end{figure}

We processed the standard data reduction techniques using the ESO's EsoRex pipeline \citep{Freudling2013}. We used the fors\_bias, fors\_img\_sky\_flat, and fors\_img\_science pipeline recipes from EsoRex to perform bias subtraction and flat fielding. Considering that the W2246 field has several bright stars, especially two heavily saturated bright stars, we used \textsc{photutils} package in \textsc{Python} to do the background subtraction. For the photometric calibration of the $R$, and $V$ bands images, the Stetson photometric fields \citep{Stetson2000} were observed in these bands. The Stetson photometric standard stars catalog was used for the calibration of the $R$, and $V$ bands images. The process was done using fors\_zeropoint and fors\_photometry pipeline recipes from EsoRex. Spectrophotometric standard stars \citep{Turnshek1990,Hamuy1992,Hamuy1994} were observed in $NB$ to calibrate the narrowband images. We used \textsc{PYPHOT}\footnote{\url{https://mfouesneau.github.io/pyphot/index.html}} to perform the photometric calibration. The calibrated frames are stacked to generate a deeper image for each band by using \textsc{SWarp} \citep{Bertin2002}. The FWHMs of the point spread function (PSF) are 1\farcs39, 1\farcs21, and 0\farcs97 in the $NB$, $R$, and $V$ bands, respectively. In order to measure the colors within the same aperture, we match the PSF of the $R$ and $V$ bands images to the $NB$ image (the one with the worst PSF) using \textsc{photutils}. Photometry was performed using \textsc{SExtractor} \citep{Bertin1996} in dual-image mode with the $NB$ image as the detection image. We use $2\farcs5$ diameter aperture magnitudes for the color measurements. The $5\sigma$ limiting magnitudes (with $2\farcs5$ diameter aperture) of the $NB$, $R$, and $V$ bands are 25.5, 26.2, and 26.9 mag, respectively. We adopted AUTO magnitudes as the total magnitude of detected sources. The effective area of the reduced image is 33.6 arcmin$^2$. Regions that contain saturated bright stars and severe stray light caused by instrument are masked. Figure \ref{fig1} shows the $NB$ image of the W2246 field.
We estimated the detection completeness of the $NB$ image as a function of total $NB$ magnitude. We randomly distributed mock point sources (gaussian profile, FWHM = 1\farcs4) with different magnitude into the effective area of $NB$ image. Then, we ran the SExtractor with the same detection parameters, and checked whether the added source was detected or not. The results are shown in Figure \ref{fig2}. The detection completeness drops to about 80 percent at $NB_{tot} = 25.1$ mag.

\begin{figure}
\plotone{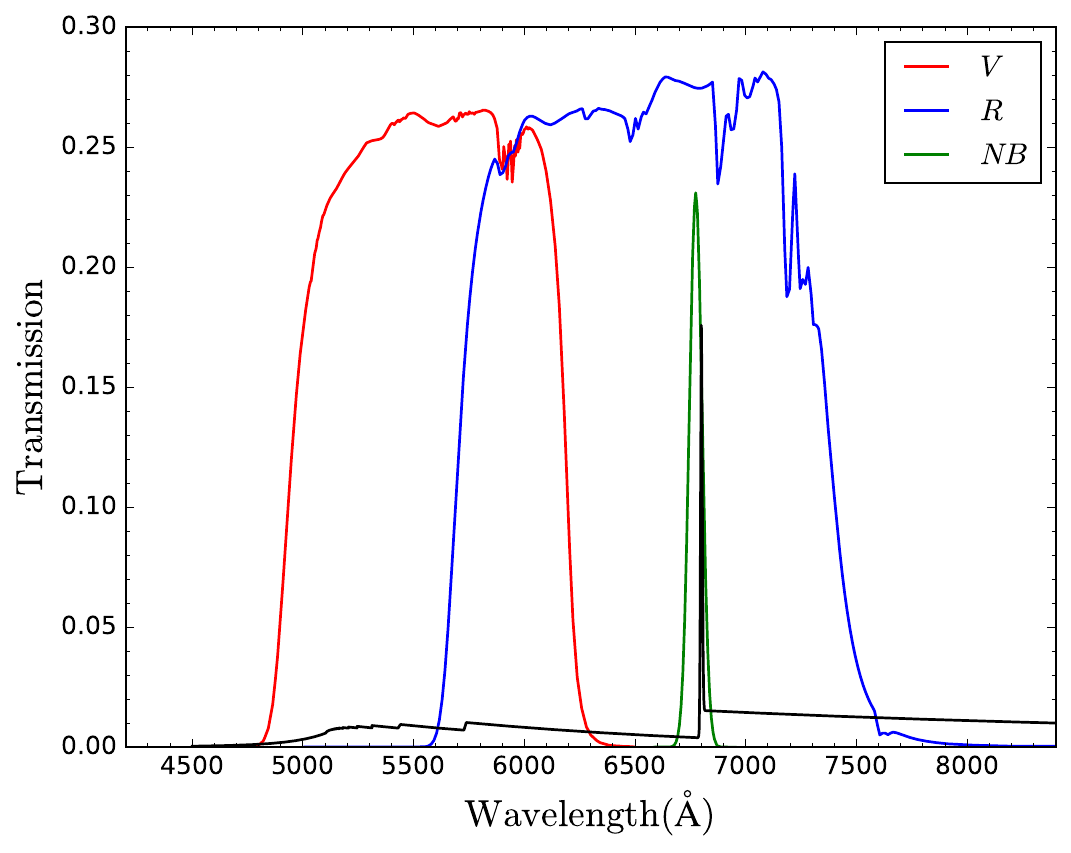}
\caption{
Total transmission curves of $NB$, $R$, and $V$ filters, including the instrument response and atmospheric transmission at airmass of 1.2. The LAE model spectrum corresponds to a galaxy with UV continuum slope $\beta=-2$ and rest frame Ly$\alpha$ equivalent width $EW_0 = 20$ \AA. Redshift of this model spectrum is $z=4.6$. The LAE model spectrum are corrected for the IGM absorption \citep{Inoue2014}.
\label{fig3}}
\end{figure}

\begin{figure*}
\epsscale{1.2}
\plotone{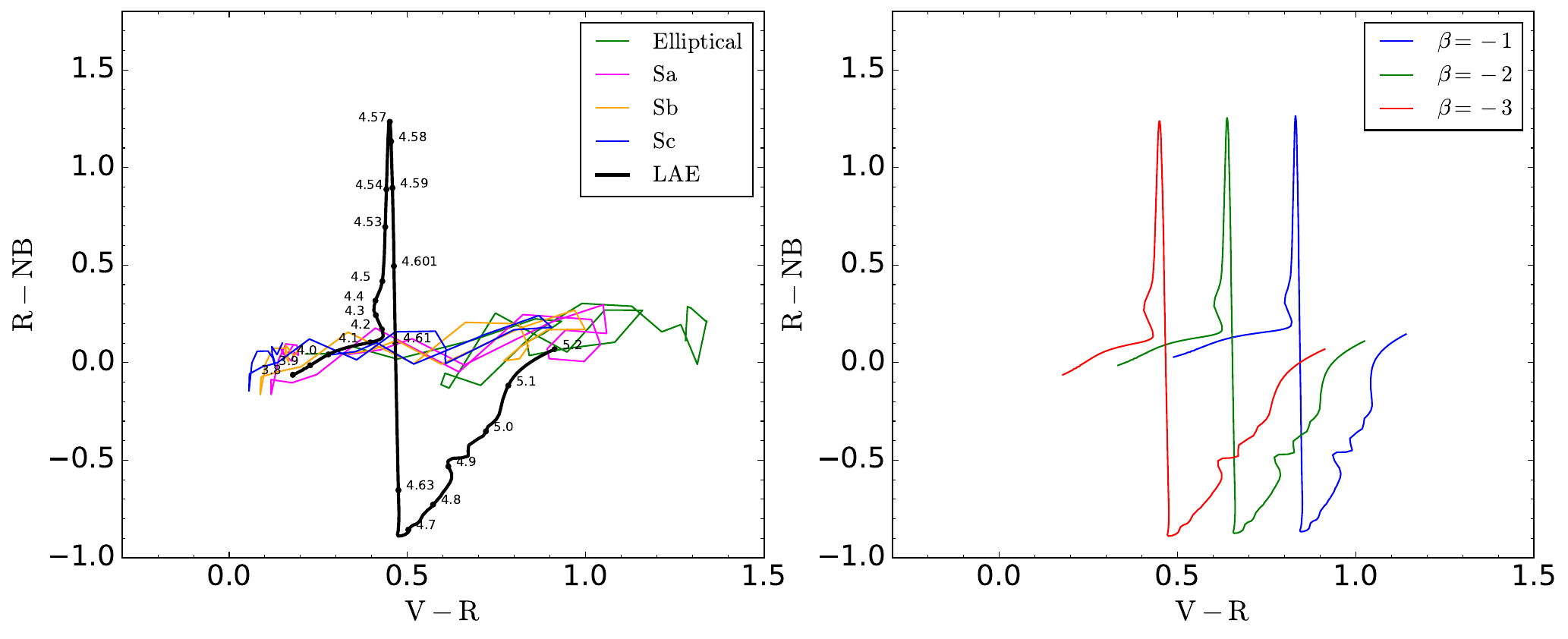}
\caption{
Left: evolutionary track of an LAE with $EW_0 = 20$ \AA (black curve). Points over the black curve indicate colors of LAEs from redshift 3.8 to 5.2, and the redshift of W2246 $z_{\rm [CII]}=4.601$ is also shown. We also plotted the evolutionary tracks of possible contaminants redshifted from 0 to 3. Green, magenta, orange, and blue curves are the evolutionary track of elliptical, Sa, Sb, and Sc galaxies, respectively. Right: we show the evolutionary track of LAE model spectrum with three different UV slopes, ranging from $\beta=-3$ to $\beta=-1$.
\label{fig4}}
\end{figure*}

\section{LAEs Selection}\label{sec:selection}

\subsection{Selection Criteria}\label{sec:criteria}

We use color selection criteria to select LAEs from photometric catalog, which is commonly used in previous studies \citep{Venemans2002,Venemans2005,Ouchi2008,Mazzucchelli2017,Garc2019,Hu2019,Liang2021}. To define the selection criteria, we assume that the LAE model spectrum at $z\sim4.6$ has a Gaussian profile Ly$\alpha$ emission with a power-law UV continuum ($f_{\lambda} \propto \lambda^{\beta}$). The LAE model spectrum are corrected for the IGM absorption with the model from \citet{Inoue2014}. Then, we convolve the model spectrum with filters total transmission curves (including the instrument response and atmospheric transmission) to calculate the magnitude in each filter. Finally, the color excess can be estimated from this simulation. The total transmission curves of three filters and the LAE model spectrum are shown in Figure \ref{fig3}. We consider the Ly$\alpha$ emission of model spectrum has a rest equivalent width $EW_0 = 20$ \AA, which is the conventional value to define LAEs \citep{Ouchi2020}. The predicted track of LAE model from redshift 3.8 to 5.2 is plotted in the color-color diagram in the left panel of Figure \ref{fig4}. Different UV continuum slopes $\beta$ from -3 to -1 are set in the simulation and are shown in the right panel of Figure \ref{fig4}. We find that when UV continuum slopes $\beta$ changes, it mainly affects the color of $V-R$, and has little effect on the color of $R-NB$.
We also consider possible contaminants such as low redshift {\sc [Oii]} emitters, {\sc [Oiii]} emitters, elliptical galaxies, and spiral galaxies (Sa, Sb, Sc) in our estimation of the color selection criteria. The number densities of {\sc [Oii]} and {\sc [Oiii]} emitters at low redshift had been studied in \citet{Pirzkal2013}. Following the calculation of \citet{Zheng2013}, we estimate the number of {\sc [Oii]} and {\sc [Oiii]} emitters are 0.03 and 0.04 in the W2246 field, which has little impact on our LAEs selection. For the elliptical galaxies and spiral galaxies, we get these galaxies templates from the SWIRE library \citep{Polletta2007}. The predicted tracks of the elliptical galaxies and spiral galaxies from redshift 0 to 3 are also plotted in the left panel of Figure \ref{fig4}. From the color-color diagram, we can find that although $R-NB = 0.5$ mag at $z=4.601$ is smaller than the peak of $R-NB$ of LAE model, which is due to the central wavelengths of the narrow-band filter shifts from the W2246 redshift derived from ALMA. We call this phenomenon as the shift of the observation window. However, the $R-NB$ of LAE at $z=4.601$ is still significant larger than the $R-NB$ of elliptical and spiral galaxies. In addition, the $V-R$ of LAE model is about 0.4 when UV continuum slopes $\beta$ down to -3. Considering the contamination of low-redshift {\sc [Oii]} and {\sc [Oiii]} emitters in our study is negligible, we use this value in the $V-R$ color criterion. Therefore, we select LAE using the following criteria:
\begin{align}
\begin{split}
& R - NB > 0.5, \\
& V - R > 0.4, \\
& R - NB > 2\times \sqrt{\sigma_{R}^{2} + \sigma_{NB}^{2}} , \\
& 20 < NB < {NB_{lim, 4\sigma}},
\end{split}
\end{align}
We set the $2\sigma$ significant of the narrow-band excess criterion to avoid contamination by sources that satisfy the color criteria only due to photometric errors. The lower limit for the $NB$ magnitude is to avoid saturation, and the upper limit is to search for more faint LAEs while ensuring high reliability.

\subsection{LAE Sample}\label{sec:sample}

\begin{figure*}
\plotone{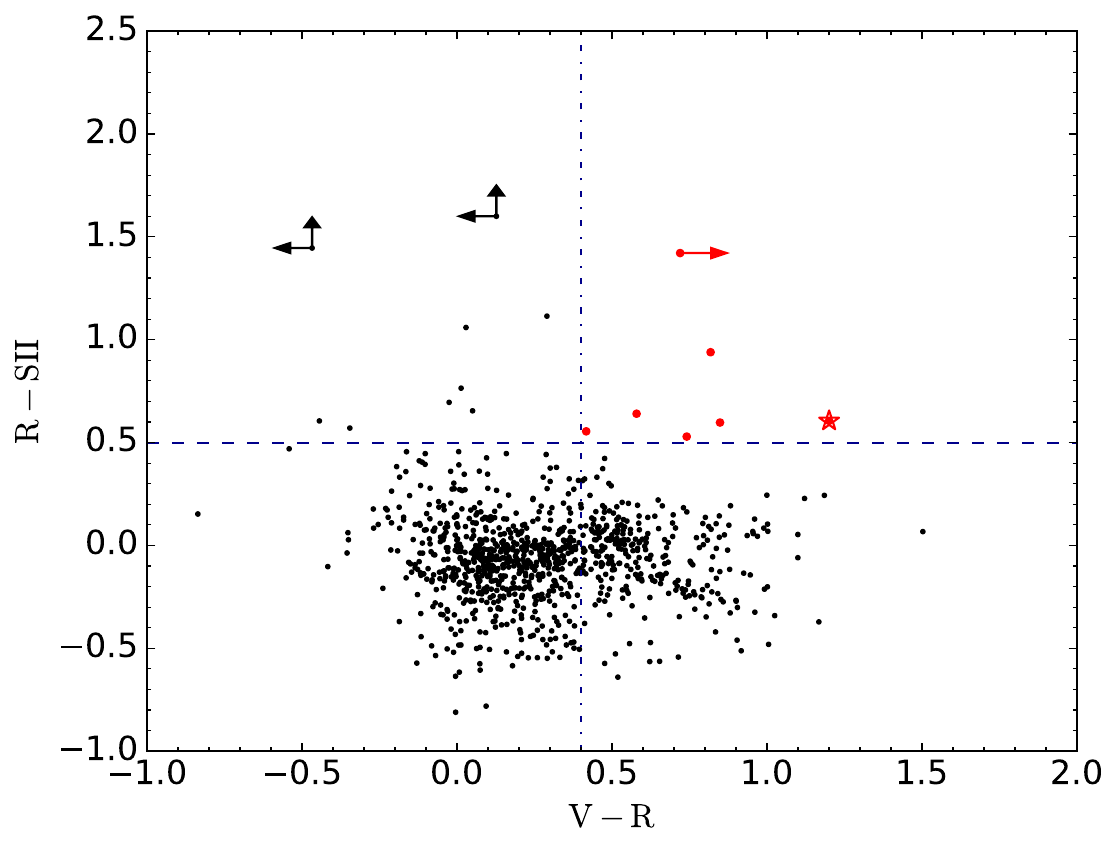}
\caption{
Color-color diagram of the sources in the W2246 field. Horizontal and vertical dashed lines corresponds color criteria $R - NB > 0.5$ and $V - R > 0.4$. Sources selected as LAEs are marked as red filled circles. Red arrow indicates lower limit for the $V - R$ color, in which the source is not detected at the $2\sigma$ level in $V$ image. Black 
arrows indicate sources that are detected under $2\sigma$ level in $R$ images. The open star marks the Hot DOG W2246.
\label{fig5}}
\end{figure*}

\begin{figure}
\epsscale{1.0}
\plotone{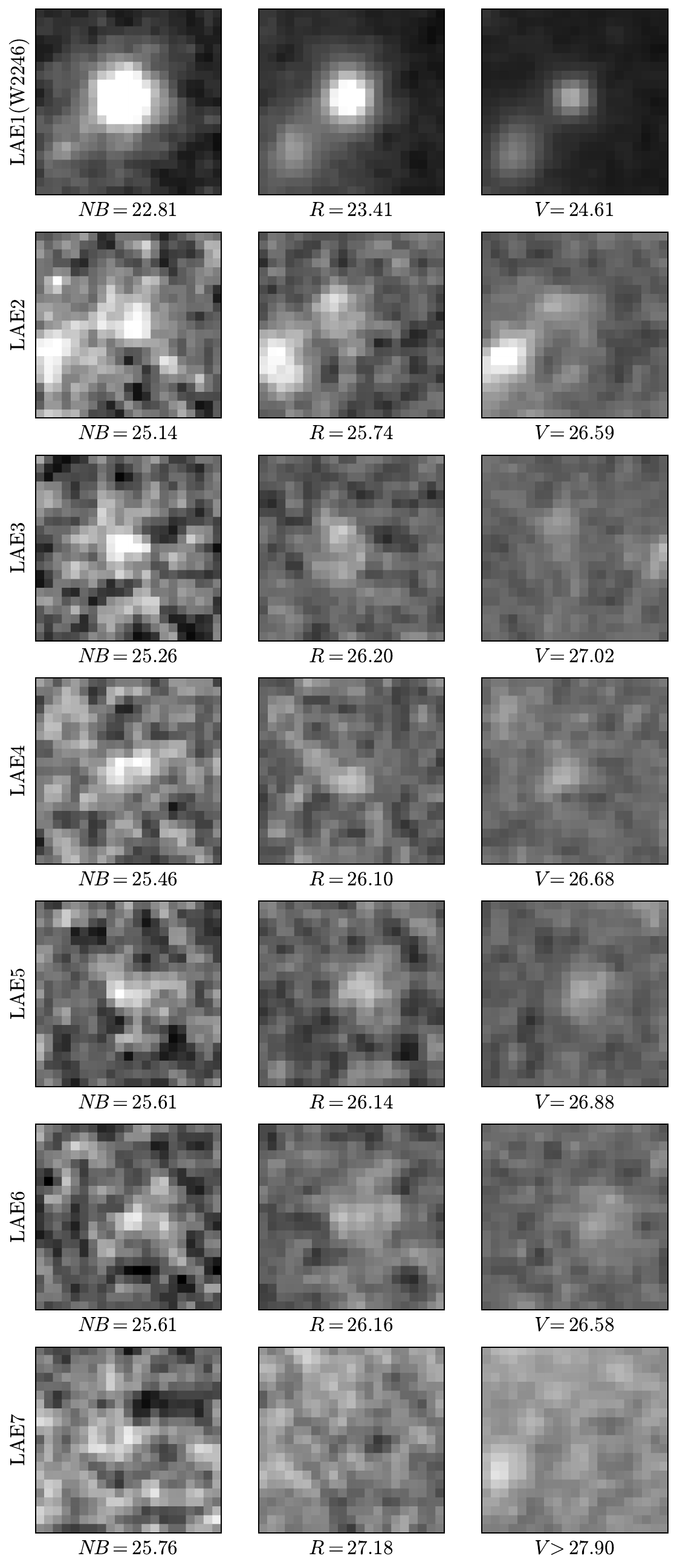}
\caption{Postage stamps of LAE candidates in $NB$, $R$, and $V$ bands. The size of each image is $5.3 \arcsec \times 5.3 \arcsec$. For sources detected under the $2\sigma$ level in broadband, its magnitude in that band is denoted as $m>{m_{lim, 2\sigma}}$.
\label{fig6}}
\end{figure}

For sources in the photometric catalog generated by \textsc{SExtractor}, we require that sources have the \textsc{SExtractor} parameter \textsc{FLAGS=0} to discard unreliable sources such as blended, saturated, or truncated sources. We only consider the sources that have $20 < NB < {NB_{lim, 4\sigma}}$ in the $NB$ image. We adopt a $2\sigma$ limiting magnitudes in $R$ and $V$ bands for marginal detections in the broadband images. In order to rule out possible spurious sources, we require the sources have $\geq 2\sigma$ detection in at least one broadband image. We show the $R-NB$ and $V-R$ color-color diagram of all the sources that satisfy the above conditions in Figure \ref{fig5}. There are seven sources that satisfy our LAE selection criteria and are considered to be LAE candidates. We show their postage stamps in $NB$, $R$, and $V$ bands in Figure \ref{fig6}. Six of the seven sources are detected in all three bands, only one source are detected in $NB$ and $R$ bands but not detected in the $V$ band. Among these LAE candidates, the Hot DOG W2246 itself is selected to be a LAE candidate. Considering W2246 was spectroscopically confirmed to be a LAE in \citet{Diaz2021}, W2246 is selected by the criterion is an expected result, and indicate the parameters for our criteria are appropriate.

\section{RESULTS}\label{sec:results}

\begin{deluxetable*}{cccccc}
\tablecaption{Photometric Properties 
\label{table:Photometric Properties}}
\tablehead{
\colhead{ID} & \colhead{R.A.} & \colhead{Decl.} & \colhead{$NB_{tot}$} & \colhead{$L_{Ly\alpha}$}  & \colhead{SFR} \\
\colhead{} & \colhead{(J2000)} & \colhead{(J2000)} & \colhead{(AB mag)} & \colhead{($\times 10^{42} \ erg \ s^{-1}$)} & \colhead{$(M_{\odot} \ yr^{-1})$}}
\startdata
LAE1(W2246) & 341.53154167 & -5.44305556 & 22.51 & 12.8 &  11.6   \\
LAE2 & 341.52625438 & -5.43420997 & 24.46 & 2.3  &  2.1   \\
LAE3 & 341.48297707 & -5.45017422 & 24.98 & 2.7  &  2.5   \\
LAE4 & 341.52000180 & -5.44773953 & 25.18 & 1.3  &  1.2   \\
LAE5 & 341.47608618 & -5.44077960 & 25.50 & 1.0  &  0.9   \\
LAE6 & 341.57522991 & -5.45600469 & 25.54 & 0.8  &  0.7   \\
LAE7 & 341.50643274 & -5.44953175 & 25.59 & 1.7  &  1.5   \\
\enddata

\end{deluxetable*}

\begin{figure}
\plotone{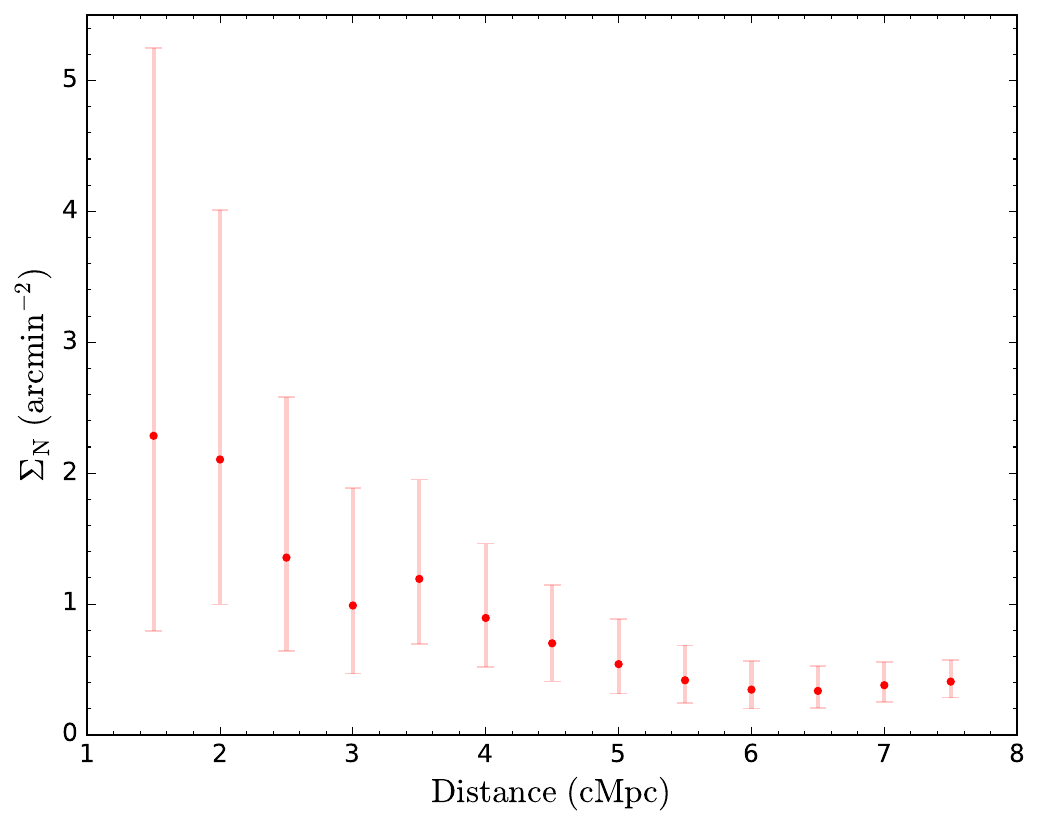}
\caption{
The surface density distribution of LAEs as a function of distance to Hot DOG W2246. The numbers of LAEs are corrected for completeness. Error bars are computed based on \citet{Gehrels1986}.
\label{fig7}}
\end{figure}

\begin{figure*}
\plotone{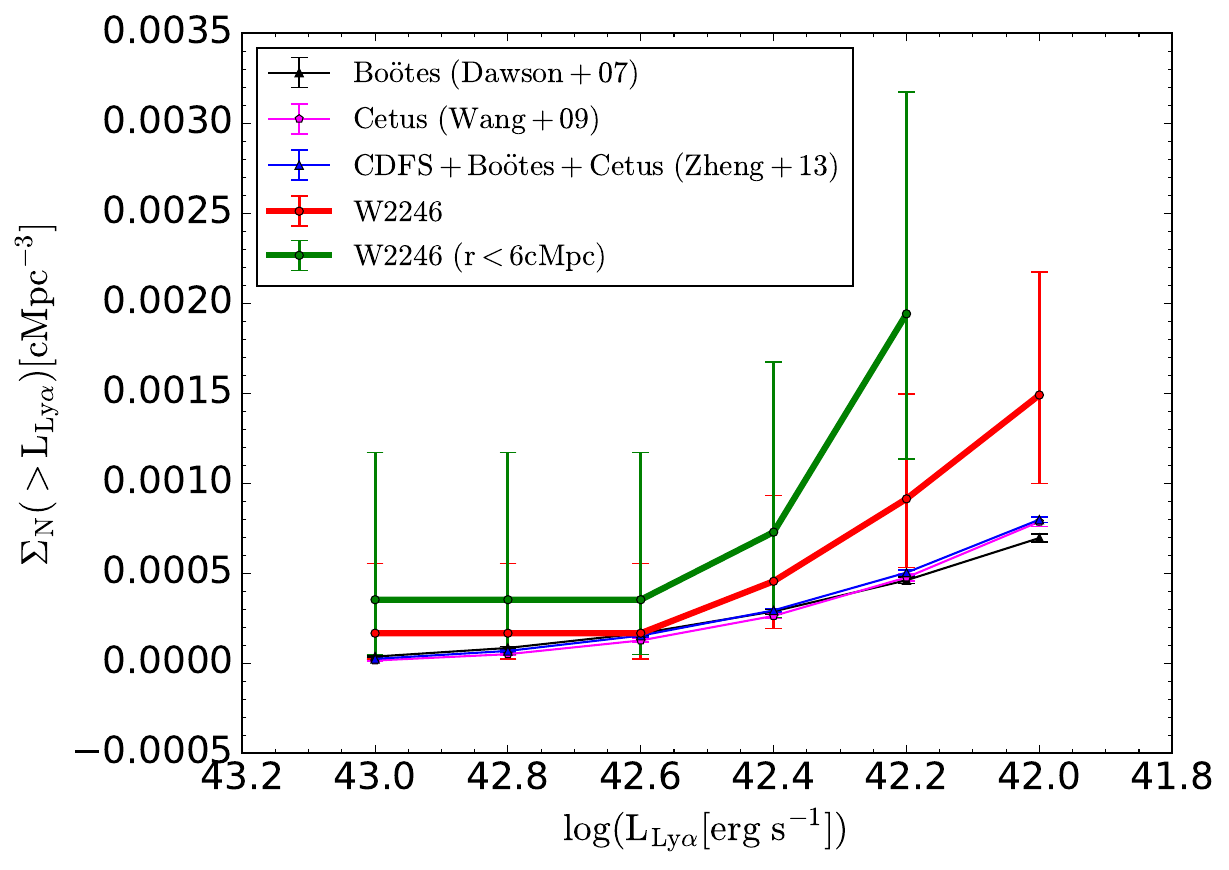}
\caption{
The cumulative number densities of LAEs in the W2246 field and the blank fields as a function of $L_{Ly\alpha}$. Error bars are based on Poisson statistics \citep{Gehrels1986}.
\label{fig8}}
\end{figure*}

\subsection{Photometric Properties of the LAEs}\label{sec:properties}
We calculate the Ly$\alpha$ line luminosity of LAE candidates from the photometry using the formulas from \citet{Venemans2005}:
\begin{align}
\begin{split}
& F_{Ly\alpha} = \frac{\Delta  \lambda_{R} \Delta \lambda_{NB}(f_{\lambda,NB}-f_{\lambda,R})}{\Delta \lambda_{R} -\Delta \lambda_{NB}}, \\
& L_{Ly\alpha} = 4\pi d_{L} ^{2} F_{Ly\alpha}, \\
\end{split}
\end{align}
In the formulas, $\Delta  \lambda_{NB}$ and $\Delta \lambda_{R}$ are the FWHM of $NB$ and $R$ filters, $f_{\lambda,NB}$ and $f_{\lambda,R}$ are the flux density in the $NB$ and $R$ filters, $d_{L}$ is the luminosity distance. We also estimate Star Formation Rate (SFR) use the following empirical relation between SFR and $L_{Ly\alpha}$ \citep{Ouchi2008}. The Ly$\alpha$  luminosity and SFR of seven LAE candidates are listed in Table \ref{table:Photometric Properties}.

\begin{equation}
{\rm SFR} (M_{\odot} \ yr^{-1} ) = \frac{L_{Ly\alpha}}{1.1\times 10^{42} \ erg \ s^{-1} }
\end{equation}

\subsection{Spatial distribution of LAEs}\label{sec:Spatial}
The spatial distribution of LAEs in the W2246 field is shown in Figure \ref{fig1}. We can intuitively find the concentration of LAEs near Hot DOGs. In order to quantitatively describe this phenomenon, we show the surface density distribution of LAEs as a function of distance to Hot DOG W2246 in Figure \ref{fig7}. The surface density of LAEs in the vicinity of Hot DOGs is higher than that of the whole field. We also find the size of the concentration region is small. When the distance increases to 6 cMpc, the surface density of LAEs rapidly drops to 1/6 of the value at the distance of 2 cMpc. The detail of analyzing the impact of size on environment study is described in Section \ref{sec:density} and Section \ref{sec:discussion}. In addition, LAEs in Figure \ref{fig1} do not seem to be randomly oriented but rather then seem to follow a filamentary-like orientation. The discussion of this phenomenon is described in Section \ref{sec:discussion}.

\subsection{Comparison with the Number Density of LAEs in Blank Fields}\label{sec:density}
In order to study the environment of the Hot DOG W2246, we compare the number density of LAEs in W2246 field to values measured in blank fields. There are several studies focus on searching LAEs at redshift $z \approx 4.5$ in blank fields such as Bo\"otes, Cetus, and CDFS fields \citep{Dawson2007,Wang2009,Zheng2013}, $Ly\alpha$ luminosity functions (LFs) are calculated based on those LAEs surveys. In particular, \citet{Zheng2013} derive a unified $Ly\alpha$ LF at $z \approx 4.5$ by combining all the LAEs found in Bo\"otes, Cetus, and CDF-S fields. According to the FWHM of $NB$ filter and the FOV of VLT/FORS2, the redshift volume $V$ of our study can be derived. Then we can calculated the expected number of LAEs in the blank fields with the same volume $V$ using the LFs. Notice that those $Ly\alpha$ LFs are built only using the spectroscopically confirmed LAEs, and the spectroscopic confirmation success rate $f_{\rm {spec-confirm}}$ for the LAE candidates (selecting through narrowband imaging) is about $\sim 0.8$ \citep{Zheng2013}. Therefore we assume $f_{\rm {spec-confirm}} = 0.8$ for our LAE candidates (except W2246 which has been spectroscopically confirmed) in this work. The number counts of LAEs are also performed completeness correction. We compare the number density of LAEs in W2246 field with the blank fields in different luminosity bin. The cumulative number densities of LAEs in the W2246 field and the blank fields as a function of $L_{Ly\alpha}$ are shown in Figure \ref{fig8}. We find that there is an overdensity of LAEs around W2246 compared to all the blank fields. The overdensity factor at $log[L_{Ly\alpha} \ (erg \ s^{-1})]>42.0$ compared to Bo\"otes field, Cetus field and combine fields are $2.1_{-0.7}^{+1.0}$, $1.9_{-0.6}^{+0.9}$, and $1.9_{-0.6}^{+0.9}$, respectively.

From Figure \ref{fig7}, we find the concentration of LAEs in the vicinity of Hot DOG W2246 and the size of the concentration region is small. The surface density of LAEs decreases rapidly with the increase of distance to Hot DOG W2246 and finally tends to stabilize at distance larger than 6 cMpc, which is 2.64 arcmin. Therefore, distance $r < 6$ cMpc may be a better scale to estimate environment of Hot DOG W2246 than using the whole W2246 field. As can be seen in Figure \ref{fig8}, if we only consider LAEs within 6 cMpc from Hot DOG W2246, the overdensity fatcor increases significantly compared to the blank fields. The overdensity factor at $log[L_{Ly\alpha} \ (erg \ s^{-1})]>42.2$ compared to Bo\"otes field, Cetus field and combine fields are $4.2_{-1.8}^{+2.7}$, $4.1_{-1.7}^{+2.6}$, and $3.9_{-1.6}^{+2.5}$, respectively.

\section{DISCUSSION}\label{sec:discussion}
\subsection{Environments of Hot DOGs}\label{sec:environment}
Hot DOGs have been found to reside in overdense environments by previous studies. Overdensities of submillimeter galaxies (SMGs) around Hot DOGs have been found using JCMT SCUBA2 850 $\mu m$ observations. Compared to blank field SMGs survey, statistical overdensities of SMGs were revealed within 1.5 arcmin radius of Hot DOGs \citep{Jones2014,Jones2017,Fan2017}. \citet{Assef2015} statistically found overdensities of \emph{Spitzer}-selected red galaxies around Hot DOGs within 1 arcmin radius compared to random pointing in blank field. For case studies, \citet{Luo2022} found an overdensity of DRGs around a Hot DOG at $z = 2.3$ using NIR selection and the  overdensity factor was 2. \citet{Ginolfi2022} found an significant overdensity of LAEs around a $z = 3.6$ Hot DOG and the measured overdensity factor was 14 using the VLT/MUSE, which revealed that the Hot DOG lives in an extremely dense environment. For W2246, using ALMA observations, \citet{Diaz2018} found that W2246 is in a multiple merger system, with at least three spectroscopically confirmed companion galaxies within a few tens of kpc, indicating a kpc-scale overdensity around W2246. \citet{Zewdie2023} found an overdensity of LBGs around W2246, and the overdensity factor was 5.8 compared to the blank field. Considering projection effects in LBGs selection could dilute clustering signal, their result suggested that such a high overdensity factor can only be measured when W2246 resides in an extremely dense environment. Those studies suggested Hot DOGs may reside in overdense regions such as protoclusters and Hot DOGs could be the brightest cluster galaxies (BCGs) \citet{Assef2015,Zewdie2023}. 

In our work, we find that there is an overdensity of LAEs around W2246 compared to the blank fields, the overdensity factor is about 2 for the whole W2246 field and about 4 for region within 6 cMpc from Hot DOG W2246. Previous environmental studies of W2246 have found a kpc-scale overdensity around W2246 \citep{Diaz2018} and an overdensity of LBGs around W2246 \citep{Zewdie2023}. Our study gives the direct evidence in Mpc scale that this most distant Hot DOG W2246 is in an overdense environment. The concentration of LAEs in the vicinity of Hot DOG W2246 suggests W2246 may reside in the densest region of the dense environment traced by itself. In addition, the impact of the shift of the observation window on our observed overdensity level needs to be considered. From the description of the narrowband filter information in Section \ref{sec:obs} and the evolutionary track of an LAE in the left panel of Figure \ref{fig4}, we can find that when the redshift of an LAE is larger than the redshift of W2246, its Ly$\alpha$ emission line will move out of the narrowband filter wavelength coverage, which makes it undetectable in our observations. Therefore, if W2246 is at the center of the overdense region, only the side of the overdense region facing us can be observed using our data. As a result, the observed overdensity level is reduced, the intrinsic overdensity level will be higher than the value we calculated. Our result is consistent with previous environments studies of Hot DOGs, suggesting that Hot DOGs may be a good tracer for overdense regions. Galaxy overdensities regions such as protoclusters at high redshift are likely unvirialized \citep{Chiang2013,Muldrew2015,Overzier2016}. Due to dense environments, mergers frequently occur during this period \citep{Overzier2016,Chapman2024}. For Hot DOGs, high merger fraction has been found in a sample of Hot DOGs \citep{Fan2016b}. More kinematics studies of Hot DOGs suggested that Hot DOG stage could be sustained by minor mergers \citep{Diaz2016,Diaz2021,Ginolfi2022}. Therefore, a scenario could be speculated that Hot DOG at the center of a protocluster merges with companion galaxies or tidal structures around, and experiences intense starbursts and rapid accretion, and final evolves to the BCG.

In Section \ref{sec:Spatial} and Section \ref{sec:density}, we suggest the region within 6 cMpc (2.64arcmin) distance from W2246 may be a better scale to estimate environment of Hot DOG W2246 than using the whole W2246 field. This suggests the size of the overdense region traced by W2246 such as protocluster could be small. Determining the scale of high-redshift protocluster is an important topic in cosmological simulations \citep{Orsi2016,Izquierdo2018}, and since the number of high-redshift protocluster found so far is still very small, our qualitative analysis of the size of the overdense region at $z=4.6$ could be useful for cosmological simulations.

The distribution of LAEs in Figure \ref{fig1} seems to follow a filamentary-like orientation. Cosmological simulations predict that galaxy formation preferentially occurs along large-scale filamentary structures in the early universe. Filamentary structures are elongated and tens of Mpc in length. Protoclusters lie at the intersection of filaments in the cosmic web \citep{Kolchin2009,Codis2012,Laigle2015,Overzier2016,Kraljic2018,Kuchner2020}. In addition, \citet{Zheng2021} and \citet{Shi2021} found that the protocluster BOSS1542 at $z=2.24$ shows a very extended filamentary structure over the scale of 23.4 cMpc. The environment study of the Hot DOG at $z=3.6$ \citep{Ginolfi2022} mentioned above also find LAEs appear to be aligned along the common orientation and shows filamentary structures. Similar structures have been found in a protocluster at $z=2.84$ \citep{Kikuta2019} and a protocluster at $z=3.09$ \citep{Umehata2019}, suggesting filamentary structure may be a general feature of protoclusters in the early universe. Combining the results of previous theoretical simulations and observational studies, we potentially suggest that the distribution of LAEs in Figure \ref{fig1} may be a filamentary structure.

\section{Summary and conclusion}\label{sec:summary}

In this work, we use VLT/FORS2 narrowband and broadband images to study the environment of the Hot DOG W2246 at $z=4.6$ on Mpc scale. We search LAEs in W2246 field using color selection criteria. We find that there is an overdensity of LAEs in W2246 field compared with the blank fields. This is the direct evidence in Mpc scale that this most distant Hot DOG is in a overdense environment. Our result is consistent with previous environments studies of Hot DOGs, indicating that the environments of Hot DOGs are overdense. The overdense environment of Hot DOGs relates to merger origin of Hot DOGs, suggesting the growth of Hot DOGs by merging may happen in unvirialized overdense environments such as protoclusters.

\begin{acknowledgments}
We thank the anonymous referee for constructive comments and suggestions. This work is supported by National Key Research and Development Program of China (2023YFA1608100). We gratefully acknowledge the support of the National Natural Science Foundation of China (NSFC, grant No. 12173037, 12233008), the CAS Project for Young Scientists in Basic Research (No. YSBR-092), the China Manned Space Project with NO. CMS-CSST-2021-A04 and NO. CMS-CSST-2021-A06, the Fundamental Research Funds for the Central Universities (WK3440000006) and Cyrus Chun Ying Tang Foundations.
\end{acknowledgments}

\vspace{5mm}
\facilities{VLT(FORS2)}

\bibliography{sample631}{}
\bibliographystyle{aasjournal}

\end{CJK*}
\end{document}